%% file: despex-paper-qapl.tex
\documentclass[submission,copyright,creativecommons]{eptcs}
\providecommand{\event}{QAPL 2014}
\usepackage{color,graphicx,times,amsmath,amssymb,url}

\input omac

\begin{document}

\title{Formal and Informal Methods for Multi-Core Design Space Exploration}

\author{Jean-Francois Kempf
\institute{{\sc verimag}\\ University of Grenoble \\ France}
\email{kempf.jf@gmail.com}
\and
Olivier Lebeltel
\institute{{\sc cnrs-verimag}\\ University of Grenoble \\ France}
\email{Olivier.Lebeltel@imag.fr}
\and
Oded Maler
\institute{{\sc cnrs-verimag}\\ University of Grenoble \\ France}
\email{Oded.Maler@imag.fr}
}

\def\titlerunning{Formal and Informal Methods for Multi-Core Design Space Exploration}
\def\authorrunning{J.-F.~Kempf, O.~Lebeltel, O.~Maler}

\maketitle

\begin{abstract}
We propose a tool-supported methodology for design-space exploration for embedded systems. It provides means to define high-level models of applications and multi-processor architectures and evaluate the performance of different deployment (mapping, scheduling) strategies while taking uncertainty into account. We argue that this extension of the scope of formal verification is important for the viability of the domain.
\end{abstract}

\section{Introduction}

Consider an application program to be executed on a \emph{multi-core platform}. The application is modeled as a \emph{task graph}, a collection of tasks partially-ordered according to precedence and annotated by execution times and data transfer volumes. We assume that these durations, as well as the arrival times of new jobs to execute, admit some \emph{bounded uncertainty}. We want to evaluate the influence of different deployment strategies (mapping, scheduling, etc.) on the overall \emph{performance} of the system. We start by explaining why this research direction constitutes a fruitful and important extension of the scope of formal verification.

\subsubsection*{Correctness vs.\ Performance}

Algorithmic formal verification is concerned with proving \emph{functional correctness} of certain systems, most notably finite-state systems such as communication protocols and digital hardware. This is often done by abstracting away from \emph{data} and focusing on \emph{control} (synchronization). However, functional correctness in the strict sense often used in verification is not a \emph{necessary} nor \emph{sufficient} condition for the usefulness of a system. A bullet-proof correct system with an extremely slow response is not likely be ever used, while systems that work well \emph{most} of the time are all around us. To keep formal verification and its insights alive and make it applicable and relevant to system design beyond the very narrow context in which it is currently used, one should rethink some of the basic premises of the field,  in particular:
\begin{enumerate}
\item The qualitative logical models of systems;
\item The qualitative \emph{yes/no} nature of the questions asked and the answers provided;
\item The \emph{universal quantification} over behaviors.
\end{enumerate}
Relaxing the first premise is of course not new. Models of automata augmented with numerical variables are used extensively in \emph{software verification} as well as in \emph{hybrid systems}. Timed automata \cite{ad94}, the model most relevant to the present paper, have been invented to model delays and execution times in a quantitative way. The second relaxation which has been argued under the banners of \emph{quantitative analysis/synthesis} \cite{henzinger-quant-emsoft,BloemCHJ09} consists of decorating transition systems with numerical costs and tracking their evolution. Such costs typically admit a simpler dynamics than more general numerical variables in programs or hybrid systems. For example, the model of linearly-priced timed automata \cite{priced,BouyerFLM11}, which are timed automata augmented with costs that can grow at different rates at different states, is simpler to analyze than other hybrid systems with constant slope \cite{integration-graphs,what-decidable,pcd} because the cost variables are \emph{passive observers} of the dynamics. The relaxation of universal quantification is what underlies \emph{statistical model-checking} \cite{younes,stat-alex,stat-larsen} and can be viewed cynically as the verification community discovering what practitioners have known all along. We argue and demonstrate in this paper that a \emph{combination} of all these relaxations has a great potential in solving real problems in modern \emph{systems design} including a central problems related to the multi-core revolution: how to evaluate and optimize the \emph{performance} of application programs on such execution platforms.

Functional correctness and good performance are complementary and sometimes conflicting evaluation criteria. In \emph{hard} real-time systems, performance is hardwired into correctness: a feedback function of a controller should be computed between every two consecutive sensor readings which puts a \emph{deadline constraint} on its computation time. Using a \emph{timed} model of the software/hardware architecture, which represents the execution times of the tasks as well as  the scheduling policy, one can verify that such a deadline is never missed. In other words, the quantitative timing information about the system participates in the proof of a functional \emph{yes/no} property. In certain simple situations studied extensively by the real-time community \cite{buttazzo-book,liu-book,kopetz-book} one can do the calculation \cite{leyland} without invoking an explicit dynamic ``executable" model at all.
For other, increasingly more popular, classes of embedded systems,  the real-time constraints are \emph{softer} and the system is expected to give a best effort performance depending on the system load and resource availability. A typical example would be video streaming where a good trade-off between response time and image quality is sought. For such systems, the actual response time is a performance measure of the system, together with additional criteria such as system price or power consumption. Unlike what is common in verification, the quantitative measures are not ``Booleanized" via predicates/constraints into a yes/no answer but remain \emph{quantitative} and can be used to \emph{compare} the relative performance of different designs.

The major contrast with the tradition of safety-critical verification is that soft systems are \emph{not} evaluated according to their \emph{worst-case} behavior but in a more \emph{probabilistic} fashion. The traditional verification approach to the problem of performance evaluation based on ``classical" timed automata technology \cite{kronos,doty,if,uppaal,red,rabbit,times} is exhaustive: it can compute performance measures such as termination time and other costs for \emph{all} possible values of the uncertainty space, thus compute lower- and upper-bounds on termination time. For soft real-time systems this is, at the same time, too much and too little. The lower and upper-bounds represent very extreme cases which are realized only when all the tasks take their extremal duration values. Under very reasonable assumptions they are less likely than termination times that admit many realizations (as $7$ is more likely than $12$ in dice).
In contrast with the exhaustive approach, in Monte-Carlo simulation the uncertainty space is finitely \emph{sampled} according to some distribution and each sampling point induces a single deterministic behavior whose performance is evaluated by (cheap) simulation. Such an approach is weaker than formal verification because it does not cover \emph{all} behaviors: it can, at most, \comment{And sometimes under questionable statistical assumptions.} put bounds on the probability of error or a deadline miss. On the other hand it is stronger as it can give an estimation of the \emph{distribution} and \emph{expected value} of the termination times, which can be much more useful for this type of applications than the very conservative bounds computed by the exhaustive approach.

\comment{Since quantities associated with costs of behaviors are maintained in the model, the questions that can be answered are not restricted to \emph{what is the probability of satisfying a  property} but can be extended to include questions such as \emph{what is the expected performance}.
The answers can sometimes be provided analytically but more often than not they are answered by statistical methods. The uncertainty space of the system is sampled, each sample point yields a deterministic simulation from which performance data is gathered and analyzed statistically.
}

The present paper is thus yet another step toward a pragmatic fusion of formal verification and performance evaluation (see also the dedicated volume \cite{BHK-book} and proceedings of some related conference \cite{qest2011,formats2011,qapl2011}) to produce a tool-supported methodology for high-level performance analysis (and eventually, synthesis) of applications programs running on multi-core architectures. In particular, this framework provides:
\begin{enumerate}
\item A formal description language for applications, hardware platforms, external environments as well as mapping and scheduling policies;
\item A translation of these objects into timed automata and employing both \emph{set-theoretic} and \emph{probabilistic} interpretation of timing uncertainty in their semantics;
\item Performance evaluation procedures based on either standard zone-based timed automata verification (when size permits) or statistical simulation.
\end{enumerate}

\subsubsection*{Industrial Context}

Platform 2012 (P2012, \cite{p2012}) is an ongoing project of ST Microelectronics (the largest European semi-conductor manufacturer) and CEA-LETI to develop a multi-core architecture to serve as an accelerator (computation fabric) for high throughput computational tasks (video processing, radio sensing, image analysis) for embedded (smart phones) and other (TV set top boxes) devices. P2012 is viewed as an alternative to GPUs as a replacement of dedicated hardware currently used for these functions. The flexibility and productivity gains of software are supposed to compensate for a tolerable degradation in performance compared to hardware. However, writing parallel software is not a trivial matter and deploying it efficiently on the multi-core platform (mapping, memory allocation, scheduling of computations and data transfers) is a hard combinatorial optimization problem with a significant variations in performance over its feasible solutions. In some sense, the multi-core revolution brings application software developers back to earlier and darker days where they had to reason about low-level architecture dependent details in order to meet performance requirements.\ft{Or one can look at it more positively as pulling developers of hardware IP upward toward the joy of high-level software development.}

The present work has been carried out within the French regional project ATHOLE (2008-2012). One of the goals of the project was to provide high-level tools to analyze (and optimize) the performance of applications on the P2012 architecture. Current performance evaluation tools used on the hardware side, at least based on our experience, work in a very \emph{low granularity}, that is, they simulate the execution of the code on the processor in a \emph{cycle-accurate} manner. This leads to very costly simulation whose extreme precision is an overkill, especially given that often this simulator is combined with much rougher models of the interconnect infra-structure. Moreover, such an analysis requires that the application is already written and that the architecture exists, at least virtually. Our work suggests a complementary approach in which:
\begin{itemize}
\item Applications are modeled at the \emph{task}, rather than \emph{instruction}, level. This means that a piece of code is modeled as a \emph{timed process} characterized by a quantitative estimations of its duration \comment{In fact, tasks are associated with a \emph{quantity of work} which can be translated to a duration based of the frequency of the processor it executes on.} and the amount of data it exchanges with other tasks. Such a description is compatible in spirit with numerous data-flow and component-based frameworks \cite{sync,sdf1,sdf2,streamit,bip} advocated for writing such applications;
    \item We model high-level performance related features of the architecture such as processor speeds, bandwidth and latency of communication mechanisms, static and dynamic power consumption of architecture elements, etc.
\item Task durations, as well as arrival rates, are modeled as admitting \emph{bounded uncertainty}, thus compensating for the lack of detail and accuracy in the application and architecture models.
\end{itemize}
As a result we provide hardware-software co-designers with a tool for rapid design-space exploration: based on profiling or past experience, the designer may decorate the application with performance numbers (intervals and distributions alike) and compare the performance figures obtained using different platforms, mapping decisions and scheduling strategies. Such procedures accelerate feasibility checks at early design stages and can be eventually integrated into the compilation and deployment chain.

The rest of the paper is organized as follows: Section~\ref{sec:desc} presents the principles of our (extendable) system description language, describes the analysis techniques (formal and statistical) supported by our tool and provides some implementation details. Section~\ref{sec:example}  demonstrates the whole approach on case study concerning power/performance tradeoff of a skeleton of an application inspired by image processing. A short discussion concludes the paper.

\section{System Modeling and Analysis}
\label{sec:desc}

Our guiding modeling principle is to abstract away as much as possible from low-level details such as the application code itself or hardware protocols and compensate for the lack of precise information by increasing the uncertainty margins and taking this uncertainty\ft{Under-determination, using the terminology of \cite{under-det}.} seriously in the analysis. There will be several types of under-determination in the durations of tasks and data transfers or their arrival rates. These can be due to various phenomenological origins: tentative ignorance in early development stages, true data-dependent variability in the algorithms or unmodeled variability in the architecture workload and physical conditions.


\subsubsection*{Applications}

\begin{figure}
\begin{center}
\scalebox{0.45}{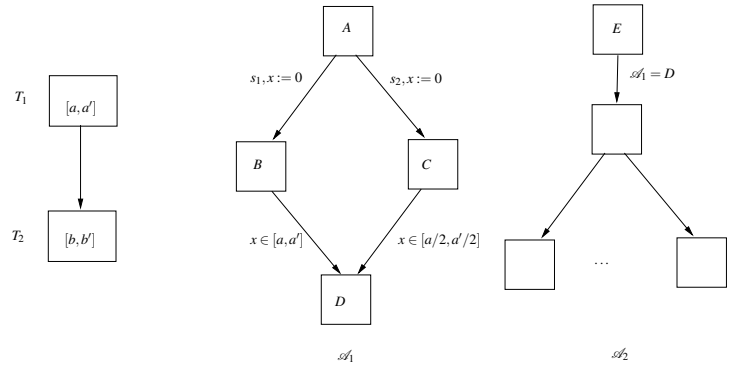 }
  \caption{Translating a task graph into timed automata.}\label{fig:task-aut-illust}
\end{center}
\end{figure}

Applications are described by task-data graphs which are a simple generalization of the common task-graph model \cite{coffman}. A task is an atomic computational entity which is characterized by an amount of work measured by instructions or cycles. Once a task is scheduled to execute on a processor with a given frequency, its amount of \emph{work} is translated into duration. We allow \emph{bounded uncertainty} (interval) for the amount of work. Another characteristic of a task is \emph{precedence}: it cannot start before some other tasks terminate and its termination may be a pre-condition to the initiation of other tasks. Finally, we model the quantity of data that has to be communicated between a task and each of its successors. Depending on the mapping of the tasks onto the architecture and the data transfer mechanism used, e.g., DMA (direct memory access) or inter-process communication, this transfer is transformed into a special \emph{communication task}. The whole task-data graph is called a \emph{job type} and it is the basic unit of work whose instances arrive to be executed. Fig.~\ref{fig:task-aut-illust} illustrates the modeling and translation to timed automata of a simple job consisting of two tasks $T_1$ and $T_2$ so that the former precedes the latter. We assume an architecture with processors that can have two speeds, $1$ and $2$. Automaton $\A_1$ models the first task. From a waiting state $A$, depending on a scheduler command, it can start executing in speed $1$ (state $B$) or speed $2$ (state $C$). In both cases the transition resets clock $x$ and as long as the automaton is in such an active state, no valid scheduler will issue another \emph{start} command for the same processor. Depending on the speed, the automaton may leave the active state when the clock is in the interval $[a,a']$ or $[a/2,a'/2]$ and move to final state $D$. The automaton $\A_2$ is similar except that it has a non-enabled state $E$ which it can leave only when $T_1$ terminates, that is, when $\A_1$ is in final state $D$. Readers are referred to \cite{schedule-tcs} for more detailed presentation of the modeling approach.

Scheduling and resource allocation in general are spread over many application domains, each focusing on specific features of the problem. The classical job-shop problem is restricted to precedence constraints that can be decomposed into a disjoint union of linear orders (jobs) and allows different types of resources (machine) so that some tasks can execute only on some machine type (which is useful for distinguishing between processors and data transfer mechanisms). On the other hand, the classical task graph problem allows a more general precedence structure (any partial order) but uses a single type of resource (processor) on which any task can be executed. Our model is a continuation of \cite{schedule-tcs} where a generalized model, fusing job-shops and task-graphs and allowing different machines as well as partial orders, was introduced and translated naturally into timed automata. We allow several job types and model the process of arrival of a \emph{stream} of job instances using input generators.

\subsubsection*{Input Generators}

Another aspect of scheduling which is treated differently along communities is the \emph{dynamic} aspect: in classical real-time scheduling new task instances arrive periodically or quasi-periodically but these are traditionally simple tasks without precedence constraints. In contrast, job-shop and task-graph problems typically do not handle the dynamic ``reactive" aspect, that is, a stream of job instances that arrive one after the other, for example, a sequence of encoded image frames or web queries. This aspect is extremely important, first because it represents the real nature of these applications and, secondly,  it favors solutions based on \emph{pipelining}, that is, the concurrent execution of tasks that belong to different job instances (see some definitions and theoretical investigations in \cite{sched-recurrent}).

One approach to treat this recurrence aspect is to use \emph{cyclic} task-graphs admitting a loop from the last to the first task. While this might be suitable for modeling loops in programs where the termination of one instance enables the execution of the next one, it is not at all natural for jobs arriving from the \emph{outside}, often independently of their processing by the system. To this end we use the concept of an \emph{input generator}, a process that generates a timed sequence of job instances subject to some logical and timing constraints. The simplest generator is the \emph{deterministic} periodic generator which produces an instance of a job every $d$ time. Strictly periodic generators are sometimes idealization of more time-noisy processes and we allow additional types of non-deterministic generators listed below (for simplicity of notation we assume here that the arrival of the first instance is $t_0=0$).
\begin{enumerate}
\item Periodic: ~~$t_k=t_{k-1}+d=(k-1)d$;
\item Periodic with jitter (non-accumulated deviations from the period):\\ ~~$t_k\in [(k-1)d,(k-1)d+J]$
\item Periodic with uncertainty:  ~~$t_k\in [t_{k-1}+d,t_{k-1}+d+J]=[(k-1)d,(k-1)(d+J)$;
\item Bounded variability: for every interval of the from $[r,r+\D]$ the number of arrival events is at most $M$;
\item Bi-bounded variability: for every such interval the number of events ranges between $m$ and $M$.
\end{enumerate}
All these types of generators are translated into timed automata that realize their semantics. They play the same role that stochastic arrival processes play in queueing theory. For generators of type $2$ and $3$ we also implemented a probabilistic semantics drawing \emph{uniformly} from $[t,t+J]$. Other types of generators that can choose (non-deterministically or randomly) among different types of jobs \cite{sched-recurrent} or stochastic generators with other distributions can be easily added.

Technically, each instance of a job generates a new instance of the corresponding automaton and this may lead to an infinite-number of automata and global states. However, we are aiming at systems that do not accumulate an unbounded backlog of unprocessed tasks and all our input generators have a finite bound on the number of instances that can arrive in a given interval of time. Thus, we can purge the automaton associated with a job when it reaches its final states and keep the number of automata which are alive in any given moment bounded. We make extensive use of the techniques developed in \cite{ramzi-thesis,BanSalahBozgaMaler09} to handle dynamic creation and deletion of timed automata, tracking the shifting denotation of clocks, etc.

\subsubsection*{Architecture}

\begin{figure}
\begin{center}
  \includegraphics[width=10cm]{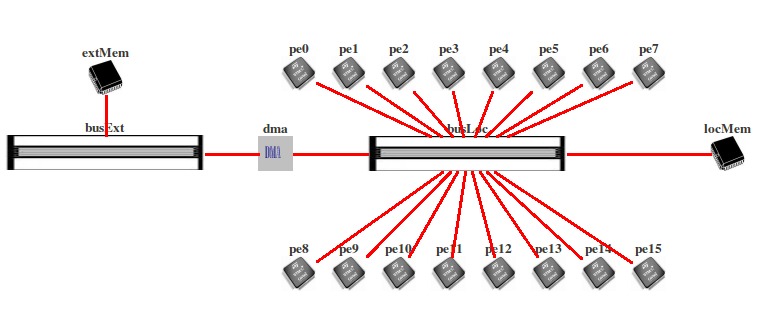}\\
\end{center}
  \caption{A model of a $16$-processor instance of P2012}\label{fig:p2012}
\end{figure}

The architecture description language (extensible as well) describes the components of the execution platform. These include:
\begin{itemize}
\item Processors, characterized by their possible speeds which may be controlled during execution and which may be turned on and off;
\item Memories defined by their access time to distinguish between slow offchip memory and fast local memory;
\item Communication mechanisms to transfer data between memories, characterized by their transfer rate, initialization costs, etc.
\item All architecture components can be decorated with power consumption figures. We assume simple system-level power models consisting of static consumption when the component is on but idle, and dynamic consumption which occurs when the components is busy (computing or moving data). Different frequencies of the processor lead, of course, to different consumption rates.
\end{itemize}
As an example, Figure~\ref{fig:p2012} shows a model, generated by the graphic user interface of our tool, associated with a $16$-core instance of the P2012 family. Each of the processors can work in several frequencies. The computation times of tasks are based on the assumption that their data resides in the local memory.\ft{To avoid confusion, note that although physically the local memory is realized in the same way as caches in more traditional processors, the programmer has full control of its contents.} The DMA agent is characterized by its initialization time. A DMA call occupies the external and internal busses for durations that depend on the amount of data and the respective transfer rates of the busses.

\subsubsection*{Deployment}
Finally, given all the above descriptions, we specify the deployment policy for the application on the platform. There are many possible types of deployment decisions and we mention some of the policies that we implemented. Adding a new policy corresponds to writing the scheduler as a timed automaton and is currently a matter of few hours, depending on the complexity of the scheduler. We have implemented a FIFO scheduler, with and without priority queues and a strict priority scheduler which may hold a low-priority task waiting although there is a free processor, to wait for a higher priority tasks not yet enabled. Each of these schedulers admits a global and a local version. In the former there is a single scheduler that may assign any task to any processing element (PE), while in the local version, the mapping of tasks to PEs is determined in advance and each PE has its own scheduler and waiting queues. A more detailed explanation of timed automata models of various schedulers appears in \cite{compare-sched}. The scheduler can also specify at which frequency to execute each task.

\comment{
\begin{enumerate}
\item FIFO scheduling: tasks that become enabled are put in a queue from which they are popped to execution on a FIFO order;and each  that becomes free pops from the queue;
\item FIFO with priority scheduling: as above but with assignment of priorities to tasks and having several queues associated with priorities and it pops the higher-priority queues first;
2.3 Strict priority scheduling: Each computation task is associated to one PE and a strict priority, each task
mapped on the same PE has a different priority. For each PE all associated
tasks will be executed according to a strict order. Each time a computation
task wants to acquire the PE it is put in a waiting set and has to wait until all
higher priority task have been executed.
2.4 Frequency scaling scheduling
Each computation task is associated to one PE, with a xed/strict priority, a
starting frequency/speed at which the PE will switch to before the beginning of
the computation and an ending frequency/speed at which the PE will switch to
at the end of the computation.
\begin{itemize}
\item FIFO scheduling: tasks are mapped to available processors and executed once they are enabled;
\item Explicit mapping and FIFO: tasks are mapped to specific processors and then
\item Explicit scheduling policy
\item
\end{itemize}
}

\subsubsection*{Analysis Methods}
\label{sec:analysis}

Once all components have been defined, their composition is equivalent to a global timed automaton whose only under-determination is related to the tasks, their durations and arrivals. We apply two types of analysis:
\begin{itemize}
\item \emph{Formal}: Using the IF toolset \cite{if} we perform on-the-fly reachability computation in the state- and clock-space. For a single job instance this type of analysis computes lower- and upper-bounds on the total termination time. For a stream of jobs, using auxiliary clocks, one can compute lower- and upper-bounds on the response time. This type of analysis manipulates timed polyhedra (zones) whose maximal dimensionality is equal to the maximal number of active system components. Moreover, with the dynamic creation and deletion of timed automata it may take more time to detect fixed-points in the reachability graph \cite{ramzi-thesis,BanSalahBozgaMaler09}. For all these reasons this type of analysis is restricted to systems that may have up to 20-25 clocks (concurrently active components).
\item \emph{Statistical}: Taking the probabilistic interpretation of temporal uncertainty we draw random values uniformly and simulate the resulting behavior. This is a fairly standard discrete-event simulation whose only particularity that it is generated based on semantically rigorous models. The runs are registered as timed traces over the alphabet of all \emph{start} and \emph{end} events. A specification in a dedicated language defines pairs of events, for example the arrival of a job and its termination, whose temporal distances are extracted from the traces. For these values we compute the mean and other statistical measures that form the basis for automatically-generated reports.
 \end{itemize}

\subsubsection*{Implementation Details}
\label{sec:impl}

The tool, temporarily dubbed \emph{The Design-Space Explorer}, consists of 25K lines of C++ code (not counting the IF analysis engine). It has a textual system description languages incorporating the abovementioned features. A graphical user interface written using Qt provides an alternative way to define systems (the illustration in this papers are produced by this interface). New types of systems components can be defined via minimal programming and are automatically propagated to the user-interface, analysis engine and the reporting system. There are two major types of outputs: raw execution traces that can be zoomed on via a dedicated \emph{trace navigator} and statistical reports in various formats. The tool architecture is summarized in Fig.~\ref{fig:tool-archi}.

\begin{figure}
\centering
  \includegraphics[width=10cm]{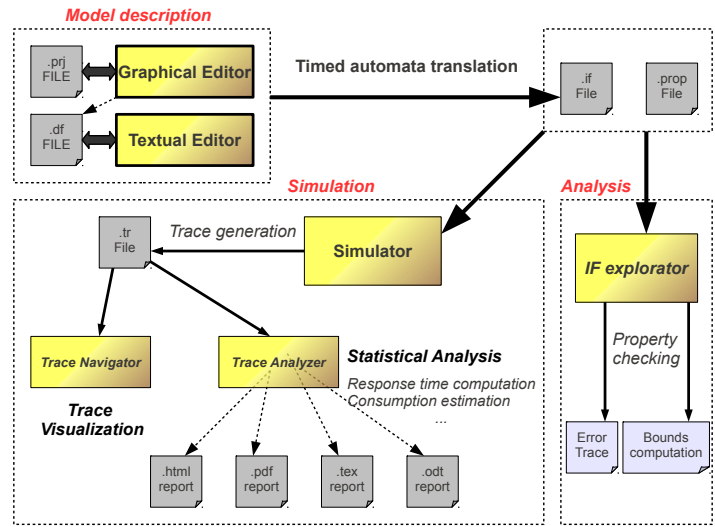}\\
  \caption{The architecture of the tool}\label{fig:tool-archi}
\end{figure}

\section{Case Study: Deploying a Video Application}
\label{sec:example}

In this section we demonstrate the applicability of our tool in exploring and comparing different deployment solutions for a data-parallel application which processes an image consisting of $16\times 16$ blocks. The image resides initially in the offchip memory and has to be brought to local memory and dispatched to the processors for execution. This is a very typical application and similar ones exist in other domains, for example in radio-sensing, the process in which a cell phone scans the bandwidth to detect channels. The sensed array of data is split into windows each undergoing the same signal processing algorithm. We will use two variants of application and of the P2012 architecture to demonstrate the functionality of our tool. All these experiments should be taken with a grain of salt concerning their realism since the development of P2012 and its applications is still in a stage where models are very approximate. The main purpose of the exploration is to illustrate the types of analysis provided by our framework.

\subsubsection*{Worst-Case vs.\ Statistics}

Consider the task-graph of Fig.~\ref{fig:app1}-(a) which represents the treatment of a horizontal band ($16$ blocks) of the image. All the blocks are fetched by a single \emph{read} command and the data is split onto $16$ tasks whose output is merged and written back to the offchip memory. Execution times for processing a single block admit up 18\% deviation from their average. We first run a TA-based analysis of the execution of this job on architecture instances with various numbers of processors to obtain the respective lower- and upper-bounds on execution times. Then we apply statistical analysis, based on $100$ random simulation runs Fig.~\ref{fig:stat1} shows a histogram of these runs for different number of processors. Note that when there is one processor per task, the average is close to the worst-case (for that configuration) because the total termination time is defined as the $\max$ of individual task termination times. On the other hand, when the number of processors is smaller and some tasks are executed sequentially, the convolution effect renders the distribution more normal-like.

\begin{figure}
\begin{center}
\begin{tabular}{ccccc}
\includegraphics[width=2.5cm]{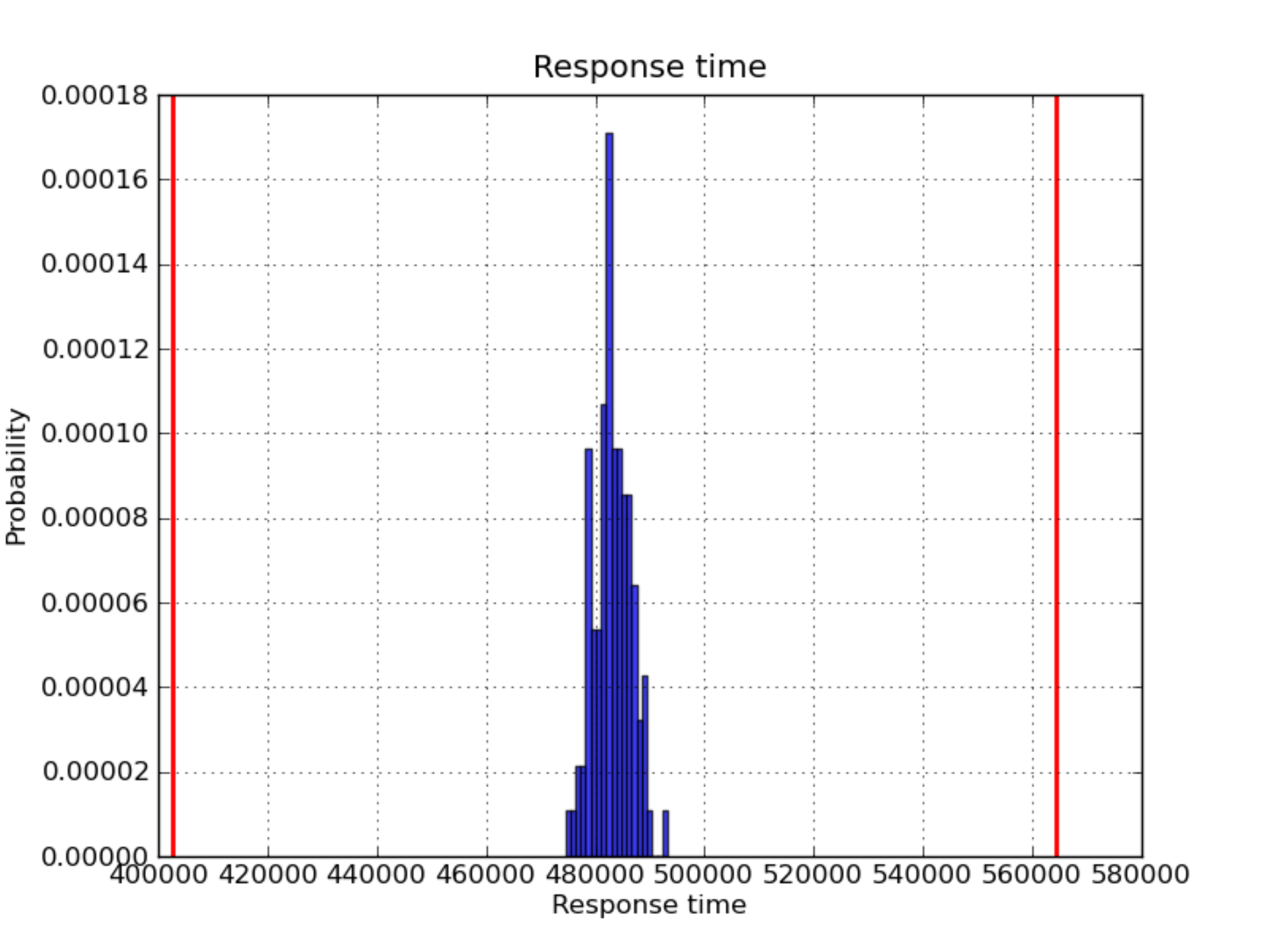} &
\includegraphics[width=2.5cm]{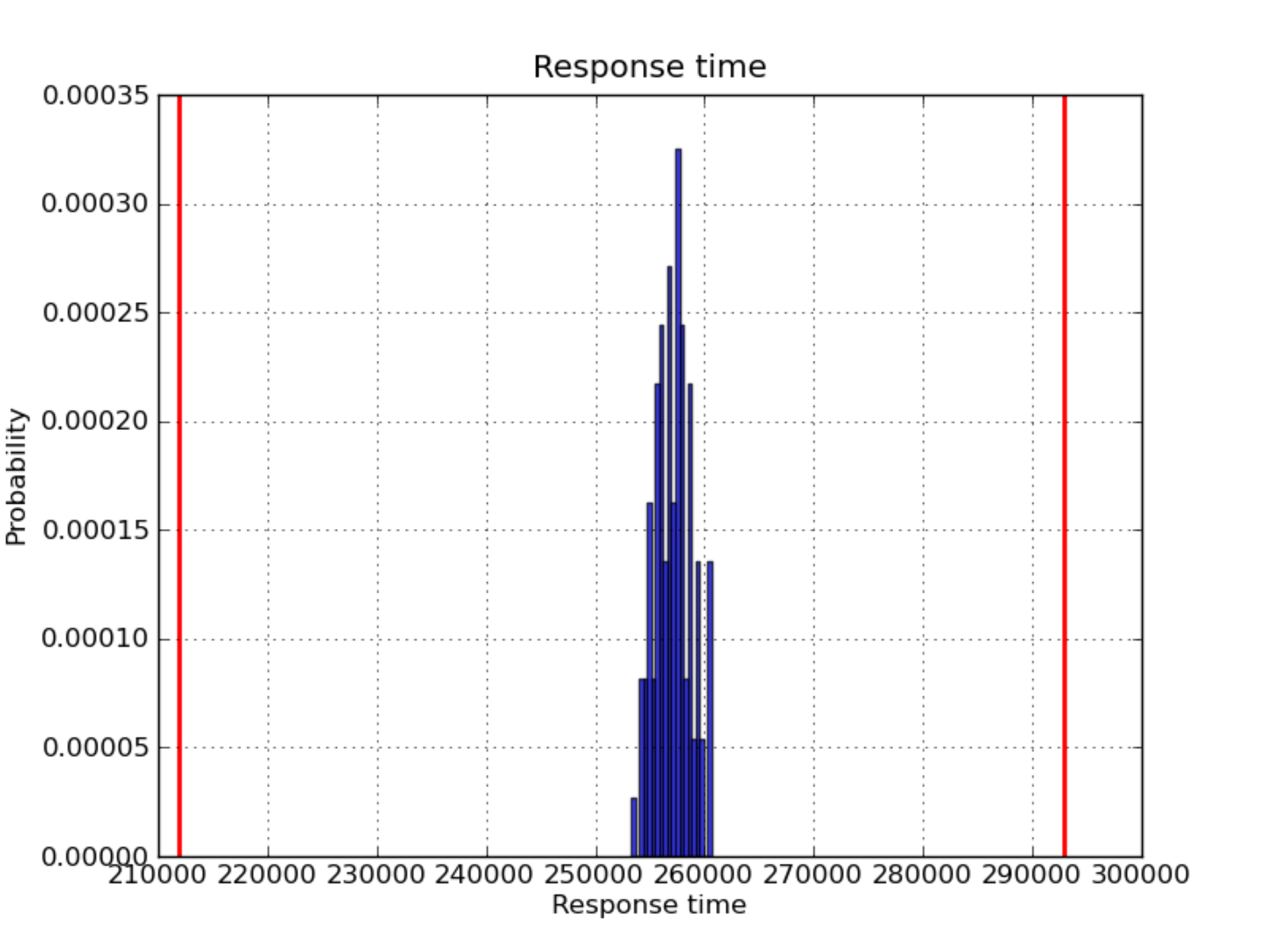} &
\includegraphics[width=2.5cm]{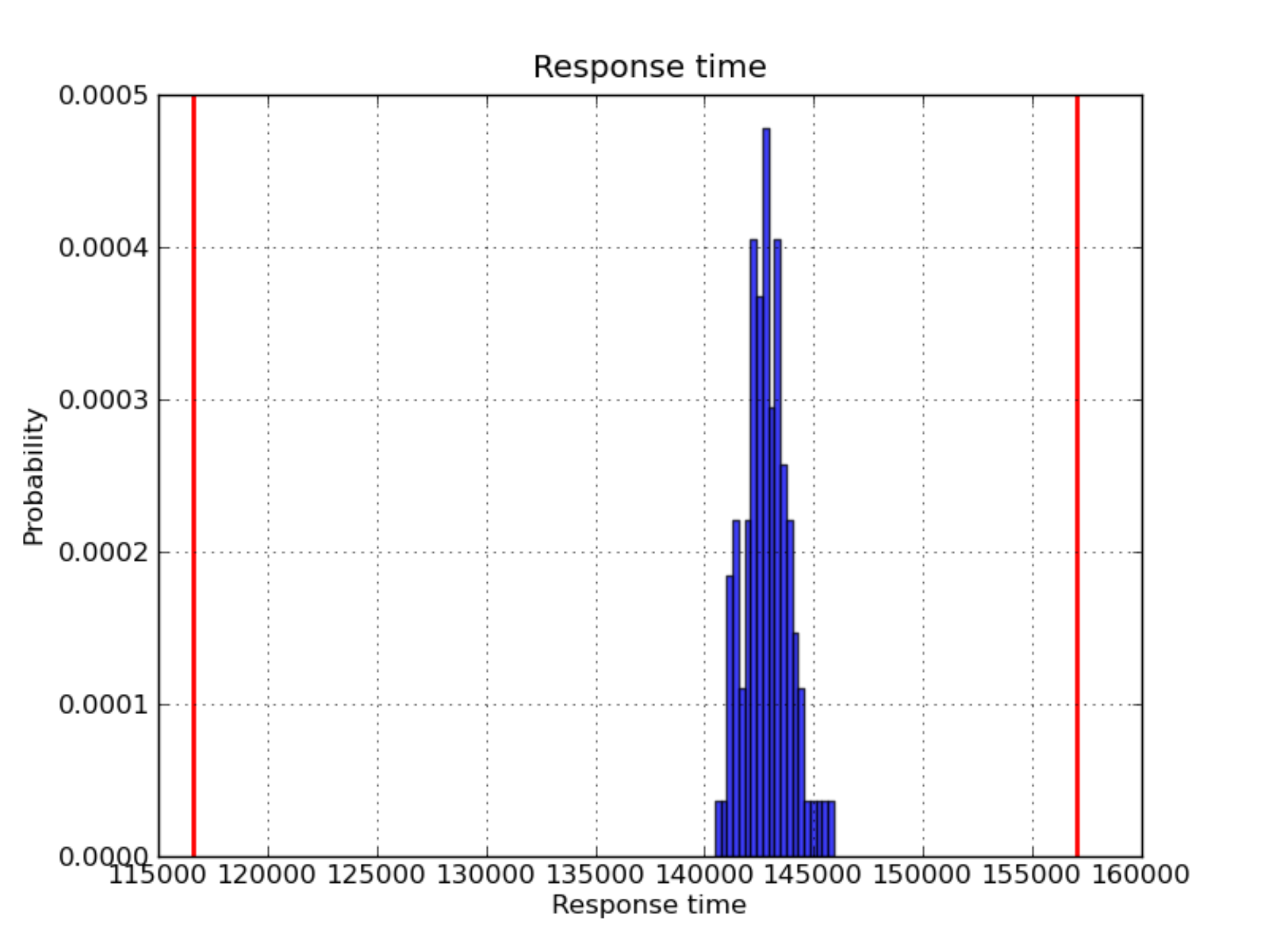} &
\includegraphics[width=2.5cm]{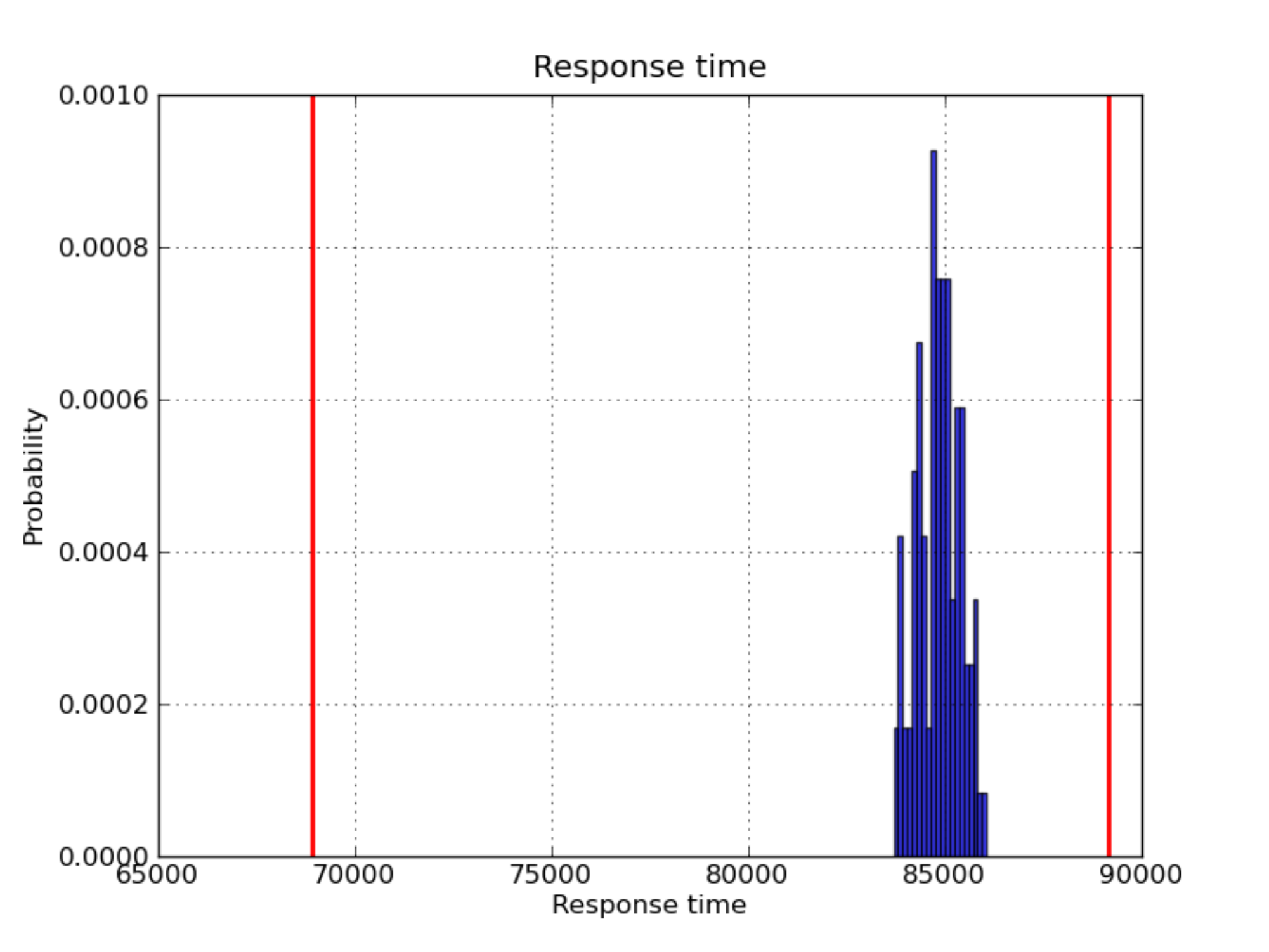} &
\includegraphics[width=2.5cm]{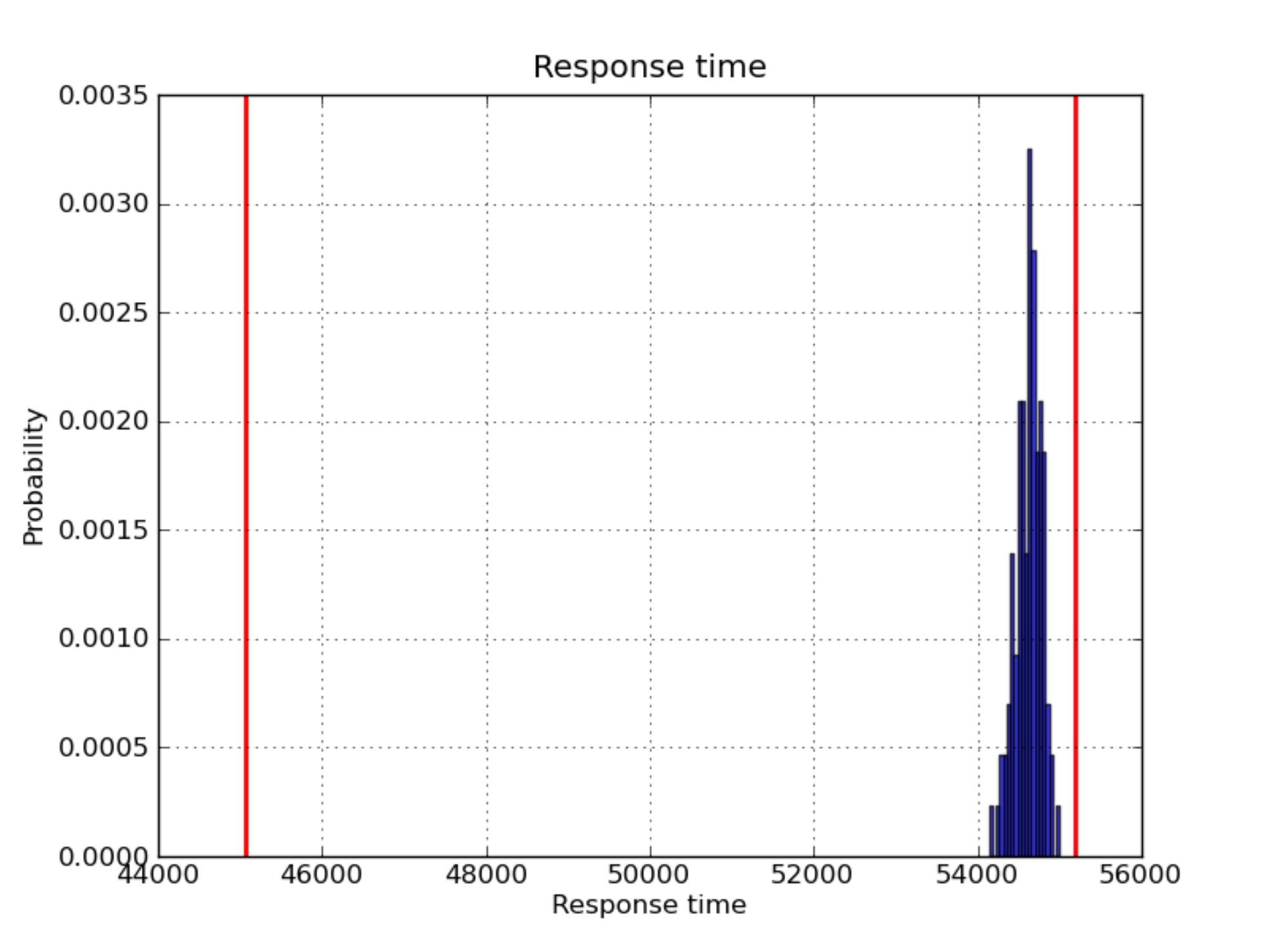}
\end{tabular}
\end{center}
\caption{The distribution of total termination times using $1$, $2$, $4$, $8$ and $16$ processors. The red vertical lines indicate the lower- and upper-bounds. Note that the vertical scaling is divided by two as we double the number of processors.}\label{fig:stat1}
\end{figure}

\subsubsection*{Reading Granularity}

We make a comparison between two strategies for fetching the data. Fig.~\ref{fig:app1}-(b) shows an alternative specification of the $16$-block computation where each block is read \emph{separately}. The whole job for $256$ blocks is represented by sequential concatenation of $16$ copies of the basic task-graphs (Fig.~\ref{fig:app1}-(a,b)). Fig.~\ref{fig:speedup-read} shows the speedup obtained by the second, more flexible policy, as the number of processors grow. Note that the speed-up in the average-case is much more significant.

\begin{figure}
\begin{center}
\begin{tabular}{c}
\includegraphics[width=9cm]{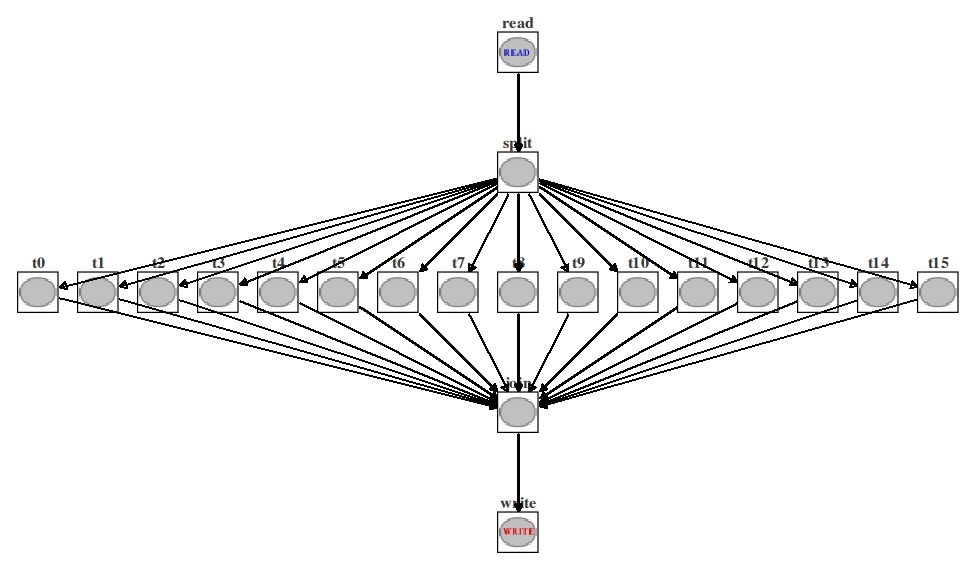}  \\ ~\\ (a) \\
\includegraphics[width=9cm]{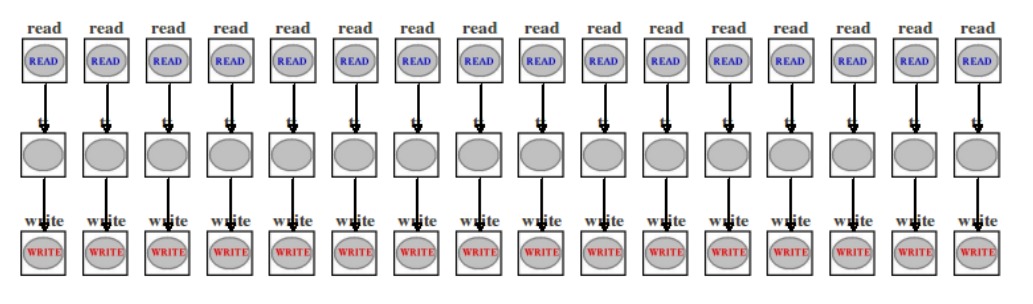}\\ ~\\  (b)
\end{tabular}
\end{center}
\caption{Two ways to process $16$ blocks of data: (a) one centralized read, split and merge; (b) $16$ independent \emph{read}s and \emph{write}s.}
\label{fig:app1}
\end{figure}

\comment{These two versions are very similar in their worst-case and best-case behaviors but vary slightly in the average case. Fig.~\ref{fig:read-compare} shows traces of the two strategies.}

\begin{figure}
\centering
  \includegraphics[width=8cm]{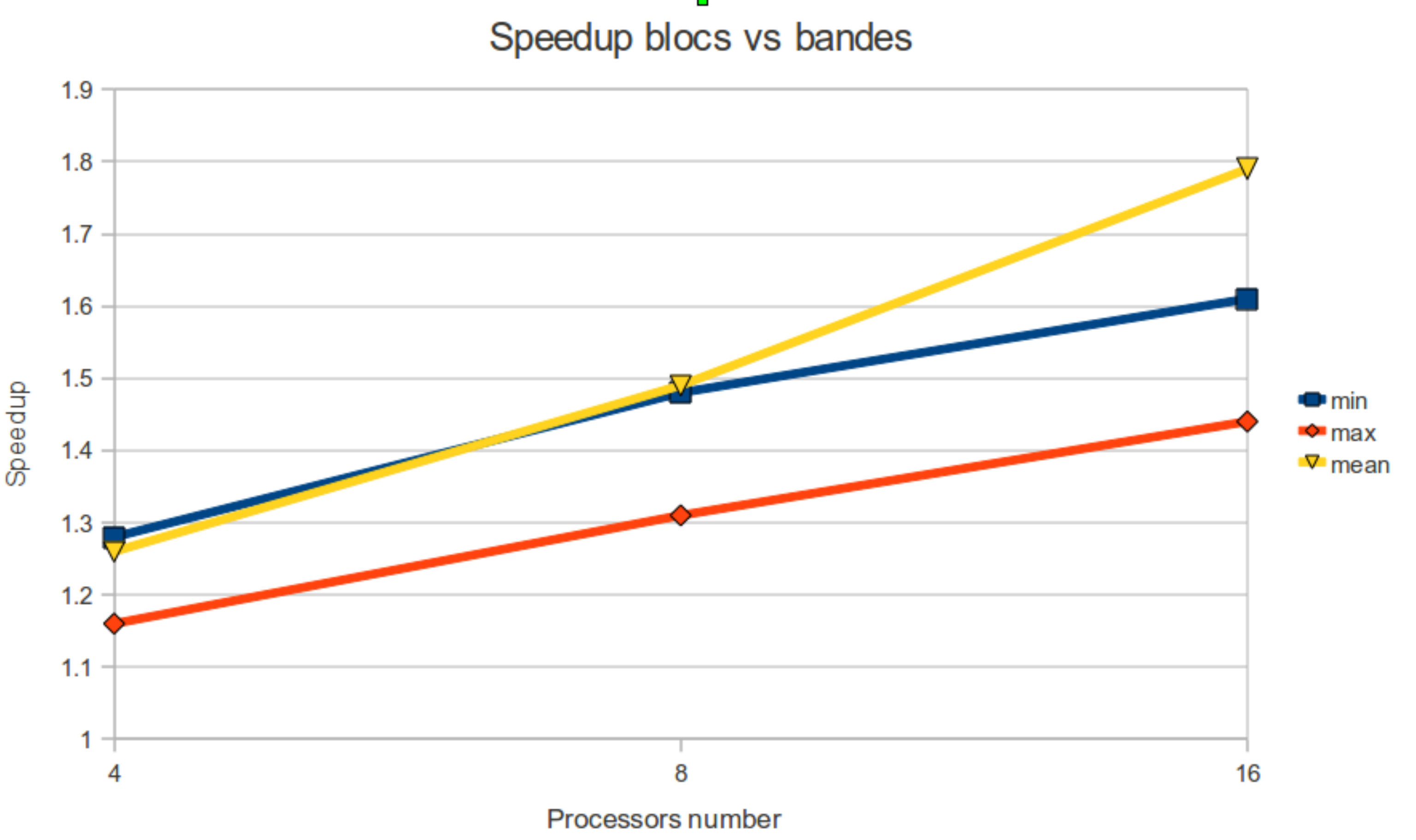}\\
  \caption{The speed-up obtained by reading single blocks compared to reading $16$-block bands.}\label{fig:speedup-read}
\end{figure}

\subsubsection*{Fixed vs.\ Flexible Mapping}

Next we move to a situation where there is a very large variability in the execution time of the tasks, namely $[150,2100]$, and compare a fixed mapping with a local FIFO scheduler for each PE against a flexible mapping by a global scheduler on an instance of P2012 with $4$ processors. We take the task graph of Fig.~\ref{fig:app1}-(b) and use a periodic event generators with jitter. Using $4$ processors, each PE is assigned $4$ tasks (exactly for the fixed mapping policy and approximately for the flexible policy) and hence the worst-case execution time for a job instance is around $8400$. For arrival periods which are smaller than the worst-case execution time, a worst-case analysis naturally shows the possibility of an unbounded accumulated backlog and, hence, unbounded latency. We perform simulations with arrival periods $7000$, $6000$, $5000$, and $4500$. Not surprisingly, the global strategy yields a much better average performance and its advantage increases with the arrival rate. Decreasing the period to $4000$ (below the average execution time) leads to frequent overflows.  Fig.~\ref{fig:flex-rigid-trace} illustrates the processor occupancy patterns following the two strategies and Fig~\ref{fig:flex-compare} shows how the relative advantage of the flexible mapping strategy depends on the arrival period.

\comment{The corner cases (best and worst) yield an identical performance for the two strategies due to symmetry}

\begin{figure}
\begin{center}
\begin{tabular}{c}
\includegraphics[width=12cm]{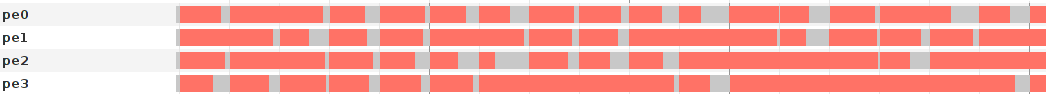} \\ ~\\ ~\\
\includegraphics[width=12cm]{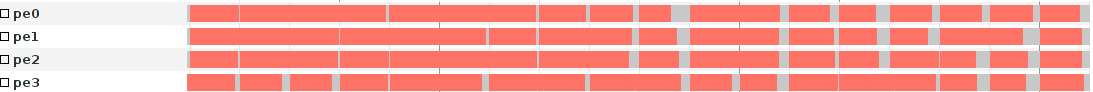} \\
\end{tabular}
\end{center}
\caption{Processor utilization under the fixed (up) and the flexible mapping strategies.}\label{fig:flex-rigid-trace}
\end{figure}

\begin{figure}
\centering
  \includegraphics[width=8cm]{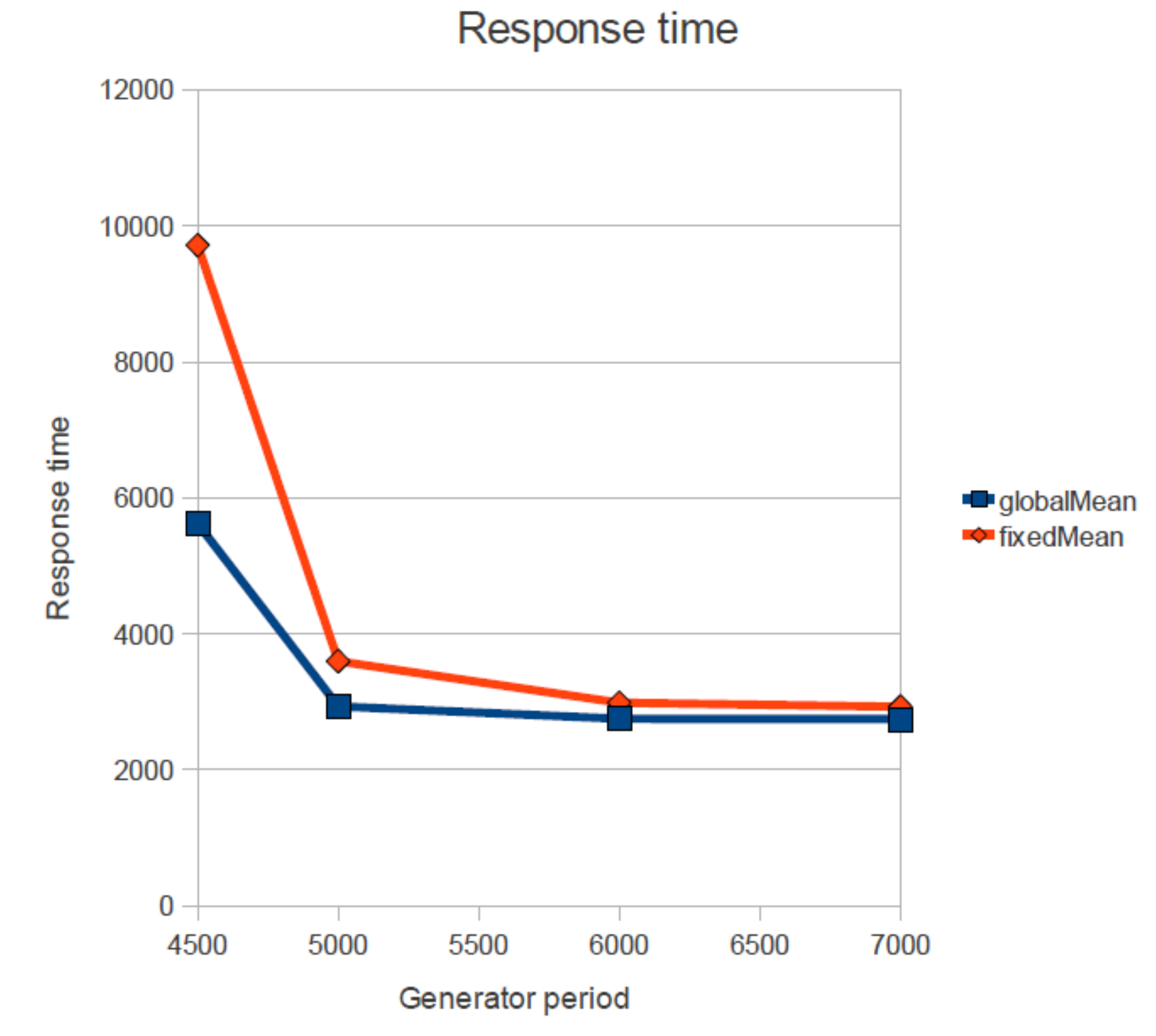}\\
  \caption{Comparing the average performance of the fixed and  flexible mapping strategies as a function of the arrival period.}\label{fig:flex-compare}
\end{figure}

\subsubsection*{Power Consumption}

In the last experiment we compare different configurations of P2012 for the trade-offs between \emph{response time} and \emph{power consumption} that they provide. We consider again a job consisting of a concatenation of $16$ copies of the task graph of Fig.~\ref{fig:app1}-(b) and execute it on instances of the architecture with $1$, $2$, $4$, $8$ and $16$ active processors, all running in either $200$, $400$ or $600$ MHz. For each configuration we run $100$ simulations and compute the average response-time and consumption. Fig.~\ref{fig:power} shows the trade-offs obtained. \comment{One can see, for example, that using all the $16$ processors at the highest frequency we can terminate very quickly but can get burnt from the generated heat (750W).} Such plots are extremely useful for detecting regions where power consumption can be significantly reduced with a modest performance degradation which still meets the system requirements. \comment{For example, moving from X to Y processors at frequency Z reduces power consumption by X\% while increasing response time by Y\.}

\begin{figure}
\begin{center}
\includegraphics[width=10cm]{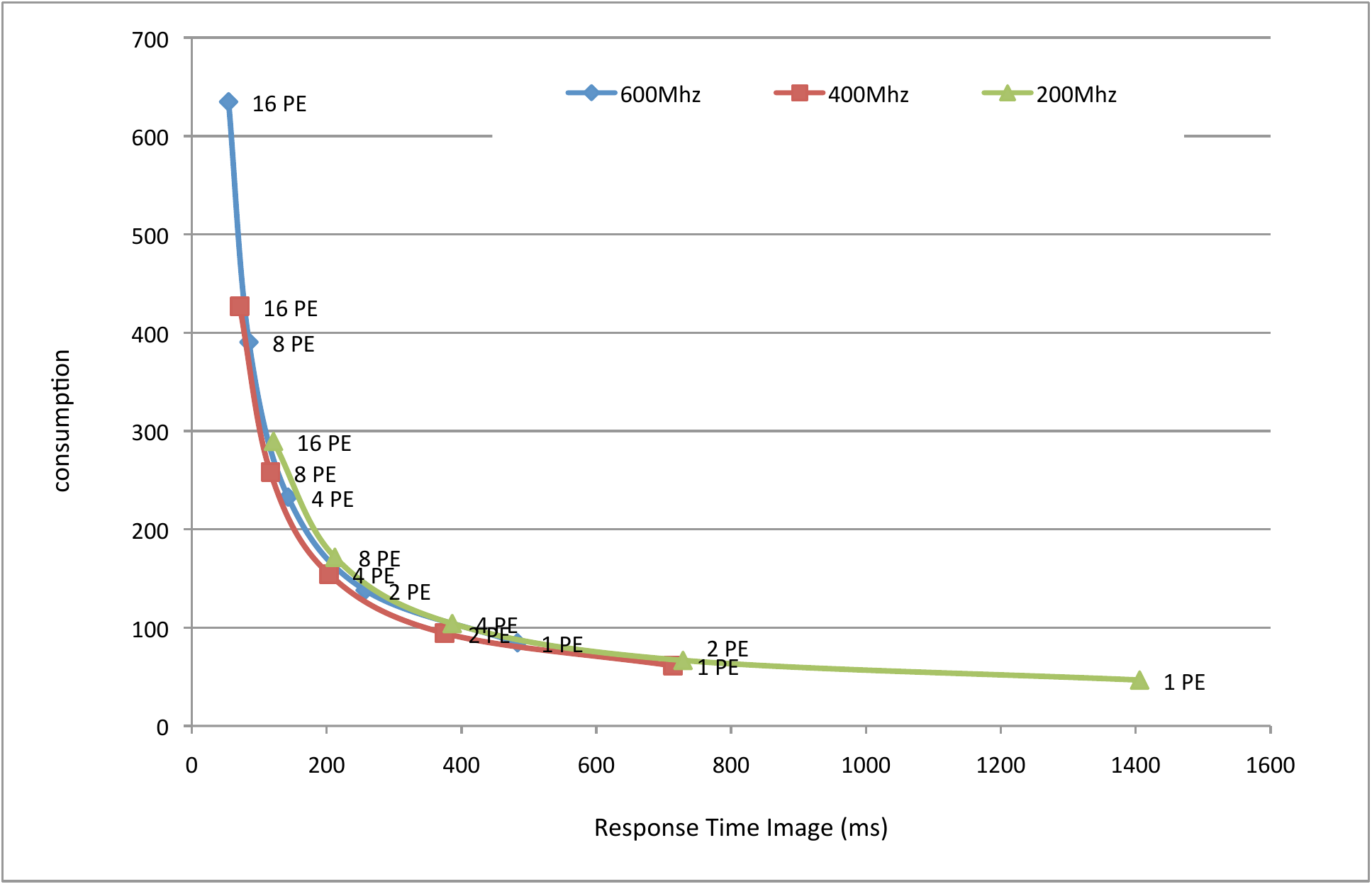}\\
\end{center}
  \caption{Power-performance tradeoffs obtained on different configurations (number of processors and frequencies).}\label{fig:power}
\end{figure}

\comment{
\subsubsection*{Why Timed Automata are Useful}
The last example shows why the analysis based on TA cannot be reduced to simple calculation of corner cases.
}

\section{Discussion}
\label{sec:discussion}

We presented what we believe to be a convincing demonstration of the potential contribution of a relaxed variant of formal verification to system design. We took from formal verification the following: 1) High-level abstract models that suppress details and focus on the features important for the task in question (traditionally synchronization and concurrency and here timing); 2) Executable and semantically correct models; 3) An explicit treatment of under-determination and 4) Tool support. We augment exhaustive timed automata analysis with Monte-Carlo simulation for scalability (similar to \cite{stat-larsen} and thus can add costs such as power consumption to the model without worrying too much about decidability. Design-space exploration is a very active topic in other communities handling embedded systems \cite{gries,teich-thiele,pimental} and we hope that our approach will contribute to bridging the gap between communities and easing the transition to multi-core computing. The main message that we want to convey is that \emph{timed models} such as timed automata are exactly the kind of models needed for this type of applications. What prevents their real-life application is their association with an overly-ambitious and intractable analysis method, which on the top of that is also hard to explain to practitioners. We believe that the more lightweight approach presented in this paper will change this situation.

Among the future extensions we consider we mention tighter integration with other formalisms used to write such applications such as synchronous data-flow (SDF) and its variants, adding a module for piecewise-analytic computation of expected performance as in \cite{compare-sched}, more sophisticated Monte-Carlo simulation, computing confidence bounds on the statistical results and more selective trace generation to reduce storage and increase speed. To promote applicability we also need to enrich the component library and  define a hierarchy of models of varying granularity and precision.

\comment{
\ni {\bf Note}: Auxiliary material including all the model files used in this paper as well as the generated reports can be found in \url{http://www.athole.org/IMG/dspex/}
}

\bibliographystyle{eptcs}
\bibliography{bib-dpa}

\end{document}

%% file: omac.tex
\def    \D      {\Delta}

\def    \A      {{\cal A}}

\def    \ni     {\noindent}

\def    \lcc#1  {\langle\!\langle #1 \rangle\!\rangle}
\def    \rcc#1  {\langle #1 \rangle}
\def    \cc#1   {[#1]}

\def    \ft#1   {\footnote{#1} }

\def	\nodeq#1 {\stackrel{#1}{\sim}}
\def	\trans#1 {\stackrel{#1}{\longrightarrow}}

\def	\comment#1 {}

\def    \comm#1  {\begin{center} {\tt $<<<$ #1 $>>>$ } \end{center}}

\def\GF{\hbox{\raise 1.6pt \hbox{$\varphi$}}}



%% file: task-aut-illust.pdftex_t
\begin{picture}(0,0)%
\includegraphics{task-aut-illust.pdf}%
\end{picture}%
%
%
\setlength{\unitlength}{3947sp}%
\begingroup\makeatletter\ifx\SetFigFont\undefined%
\gdef\SetFigFont#1#2#3#4#5{%
  \reset@font\fontsize{#1}{#2pt}%
  \fontfamily{#3}\fontseries{#4}\fontshape{#5}%
  \selectfont}%
\fi\endgroup%
\begin{picture}(10002,5044)(4156,-7020)
\put(4921,-3488){\makebox(0,0)[lb]{\smash{{\SetFigFont{12}{14.4}{\rmdefault}{\mddefault}{\updefault}{\color[rgb]{0,0,0}$[a,a']$}%
}}}}
\put(4906,-5333){\makebox(0,0)[lb]{\smash{{\SetFigFont{12}{14.4}{\rmdefault}{\mddefault}{\updefault}{\color[rgb]{0,0,0}$[b,b']$}%
}}}}
\put(4216,-3338){\makebox(0,0)[lb]{\smash{{\SetFigFont{12}{14.4}{\rmdefault}{\mddefault}{\updefault}{\color[rgb]{0,0,0}$T_1$}%
}}}}
\put(4171,-5273){\makebox(0,0)[lb]{\smash{{\SetFigFont{12}{14.4}{\rmdefault}{\mddefault}{\updefault}{\color[rgb]{0,0,0}$T_2$}%
}}}}
\put(7501,-5326){\makebox(0,0)[lb]{\smash{{\SetFigFont{12}{14.4}{\rmdefault}{\mddefault}{\updefault}{\color[rgb]{0,0,0}$x\in[a,a']$}%
}}}}
\put(9571,-5326){\makebox(0,0)[lb]{\smash{{\SetFigFont{12}{14.4}{\rmdefault}{\mddefault}{\updefault}{\color[rgb]{0,0,0}$x\in[a/2,a'/2]$}%
}}}}
\put(8686,-6196){\makebox(0,0)[lb]{\smash{{\SetFigFont{12}{14.4}{\rmdefault}{\mddefault}{\updefault}{\color[rgb]{0,0,0}$D$}%
}}}}
\put(8716,-6946){\makebox(0,0)[lb]{\smash{{\SetFigFont{12}{14.4}{\rmdefault}{\mddefault}{\updefault}{\color[rgb]{0,0,0}$\A_1$}%
}}}}
\put(12541,-6916){\makebox(0,0)[lb]{\smash{{\SetFigFont{12}{14.4}{\rmdefault}{\mddefault}{\updefault}{\color[rgb]{0,0,0}$\A_2$}%
}}}}
\put(12556,-2386){\makebox(0,0)[lb]{\smash{{\SetFigFont{12}{14.4}{\rmdefault}{\mddefault}{\updefault}{\color[rgb]{0,0,0}$E$}%
}}}}
\put(12796,-3008){\makebox(0,0)[lb]{\smash{{\SetFigFont{12}{14.4}{\rmdefault}{\mddefault}{\updefault}{\color[rgb]{0,0,0}$\A_1=D$}%
}}}}
\put(12301,-5656){\makebox(0,0)[lb]{\smash{{\SetFigFont{12}{14.4}{\rmdefault}{\mddefault}{\updefault}{\color[rgb]{0,0,0}$\cdots$}%
}}}}
\put(7501,-3061){\makebox(0,0)[lb]{\smash{{\SetFigFont{12}{14.4}{\rmdefault}{\mddefault}{\updefault}{\color[rgb]{0,0,0}$s_1,x:=0$}%
}}}}
\put(9451,-3076){\makebox(0,0)[lb]{\smash{{\SetFigFont{12}{14.4}{\rmdefault}{\mddefault}{\updefault}{\color[rgb]{0,0,0}$s_2,x:=0$}%
}}}}
\put(8791,-2371){\makebox(0,0)[lb]{\smash{{\SetFigFont{12}{14.4}{\rmdefault}{\mddefault}{\updefault}{\color[rgb]{0,0,0}$A$}%
}}}}
\put(7546,-4276){\makebox(0,0)[lb]{\smash{{\SetFigFont{12}{14.4}{\rmdefault}{\mddefault}{\updefault}{\color[rgb]{0,0,0}$B$}%
}}}}
\put(9901,-4276){\makebox(0,0)[lb]{\smash{{\SetFigFont{12}{14.4}{\rmdefault}{\mddefault}{\updefault}{\color[rgb]{0,0,0}$C$}%
}}}}
\end{picture}%